\documentclass[a4paper,5p]{elsarticle}

\journal{Renewable Energy}
\usepackage{graphicx}
\begin{document}

\begin{frontmatter}
\title{Aerodynamics of a rigid curved kite wing}
\author[a]{G.Maneia}
\author[b]{C.Tribuzi}
\author[c]{D.Tordella}
\author[c]{M.Iovieno\corref{cor1}}
\address[a]{Sequoia Automation Srl, via XXV Aprile 8, 10023 Chieri (TO), Italy}
\address[b]{Nova Analysis Snc, via Biella 72, 10098 Rivoli (TO), Italy}
\address[c]{Dipartimento di Ingegneria Meccanica e Aerospaziale, Politecnico di Torino,\\ corso Duca degli Abruzzi 24, 10129 Torino, Italy}
\cortext[cor1]{Corresponding author. Email: {\tt michele.iovieno@polito.it}}
% \author{G.Maneia,\footnote{Sequoia Automation, 10023 Chieri, Italy, AIAA Member}\\ \textit{Sequoia Automation, 10023 Chieri, Italy},\\ C.Tribuzi\footnote{Associate Researcher, Dipartimento di Ingegneria Aeronautica e Spaziale, Politecnico di Torino, Italy}, D.Tordella\footnote{Corresponding author. Associate Professor, Dipartimento di Ingegneria Aeronautica e Spaziale, Politecnico di Torino, Italy}\\ \textit{Politecnico di Torino, 10129, Italy}}
\date{-}

\begin{abstract}
A preliminary numerical study on the aerodynamics of a kite wing for high altitude wind power generators is proposed.
% A preliminary numerical study on the aerodynamics and  the flow past a curved rigid wing that is similar in shape to a tethered kite wing is proposed.
Tethered kites are a key element of an innovative wind energy technology, which aims to capture energy from the wind at higher altitudes than conventional wind towers. %by means of controlled tethered kites. %In particular, proposals to capture the wind energy by means of controlled tethered airfoils, that is, kites,  have  recently been advanced. 
%----------------------------------------
% A preliminary numerical study on the aerodynamics of a kite wing for high altitude wind power generators is presented.
% % A preliminary numerical study on the aerodynamics and  the flow past a curved rigid wing that is similar in shape to a tethered kite wing is proposed.
% Interest in kites has emerged in the context of a proposed innovative wind energy technology, which aims to captured energy from the wind by means of controlled tethered kites. %In particular, proposals to capture the wind energy by means of controlled tethered airfoils, that is, kites,  have  recently been advanced. 
We present the results obtained from three-dimensional finite volume numerical simulations of the steady air flow past a three-dimensional curved rectangular kite wing (aspect ratio equal to 3.2, Reynolds number equal to $3 \times 10^6$). Two angles of incidence -- a standard incidence for the flight of a tethered airfoil (6$^\circ$) and an incidence close to the stall (18$^\circ$) -- were considered.  The simulations were performed by solving the Reynolds Averaged Navier-Stokes flow model using the industrial STAR-CCM+ code. The overall aerodynamic characteristics of the kite wing were determined and compared to the aerodynamic characteristics of the flat rectangular non twisted wing with an identical aspect ratio and section (Clark Y profile).
The boundary layer of both the curved and the flat wings was considered to be turbulent throughout.
It was observed that the curvature induces only a mild deterioration of the aerodynamics properties. Pressure distributions around different sections along the span are also  presented, together with isolines of the average pressure and kinetic energy fields at a few sections across the wing and the wake.
% The curved and the flat wing flow fields are similar in behavior. The similarity is greater for the higher angle of attack.
Our results indicate that the curvature induces a slower spatial decay of the vorticity in the wake, and in particular, inside the wing tip vortices.

%In order to verify the correct setting and flow features of the STAR - CCM+ code,  a comparison between the numerically obtained  aerodynamic characteristics of two archetype two-dimensional  airfoil problems -- the Naca0012 and Clark Y profiles --  and  relevant experimental data in literature is also shown.
\end{abstract}

\begin{keyword}
kite \sep wind power \sep aerodynamics \sep curved wing \sep numerical simulation
\end{keyword}

\end{frontmatter}

{\section{Introduction}}
Kites have been flown in the sky for several centuries. Nowadays, they are no longer considered just a toy for children. The flight of a kite is a complex physical phenomenon and scientists have shown renewed interest in its dynamics and have investigated new applications or improved the existing ones. Till the recent past, many of such studies have concerned the use of kites as a tool to acquire meteorological data or as equipment for extreme sports, but a new frontier is now appearing: wind energy conversion.
Current wind technology, based on wind towers, has many limitations in terms of energy production, costs and environmental impact. In fact, wind turbines not only impact the surrounding environment with the land usage of their installation and with the noise generated by their blades, but the power they can provide is also limited by the low altitude at which operate, no more than 100-150 m above the ground (see, e.g. \cite{Milborrow}) .
% A major improvement could be expected by wind generators which operate at higher altitude where wind velocity is much higher. 
The possibility of collecting  wind energy at high altitudes could bring a major improvement in the design of next generation wind power plants.
This task can be achieved by using non-powered flight vehicles such as kites, which can provide a means to transfer wind energy from higher altitudes, between 500 and 1000 m above the ground, to a power conversion system on the ground by means of tethers (see, e.g., \cite{Breukels}).

% The possibility of collecting  wind energy at high altitudes plays a key role in the design of power plants for energy generation.
The design of such a high-altitude wind power generator requires a careful aerodynamic design of the kites, integrated with automated flight control.
The mathematical models of the power system configurations which have been proposed up to now -- e.g. the laddermill, the yo-yo and carousel configurations for wind energy extraction and the towing configuration for the propulsion of ships \cite{skysails, fagiano2012a} --- have focused on the design of non-linear predictive controllers \cite{Canale,canale2010,fagiano2012,fagiano2010,novara2011,Ilzhoer,Williams,Lyod}. These controllers aim
to maximize energy generation while preventing the airfoils from falling to the ground or tethers from tangling.
In these control models, a constant lift coefficient and a constant drag coefficient have always been considered for the kite wing. However, the ``engine'' of such a wind generator is a power kite and model-based control systems may not perform well without an accurate representation of the kite dynamics.
Therefore, in order to understand how kites can convert wind energy into electric energy the first step is to examine the aerodynamic performance of a kite.

\begin{figure*}
 \centering
    {\includegraphics[width=1.30\columnwidth]{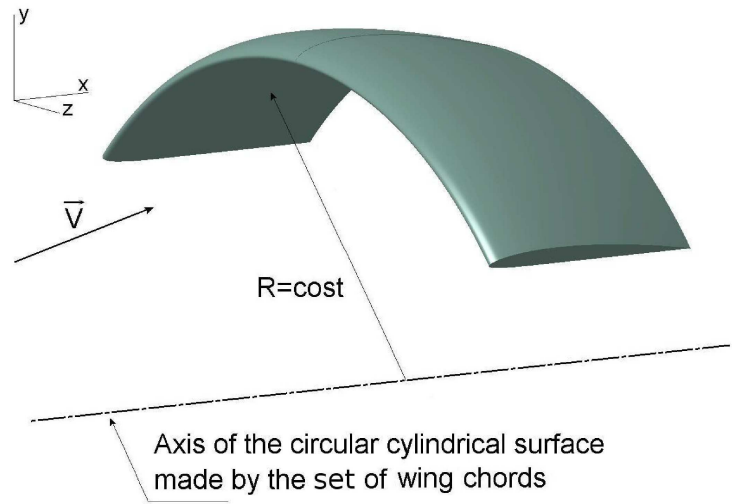}}%
\vskip -3mm
\caption{Curved wing flow schematic and reference system (the direction $x$ is the chord direction); the chords are all parallel.
 %$y$ is the direction normal to the chord and $z$ is the spanwise direction.}\\.
}
 \label{kite_scheme}
 \end{figure*}

%2
\begin{figure*}
 \centering
    {\includegraphics[width=1.5\columnwidth]{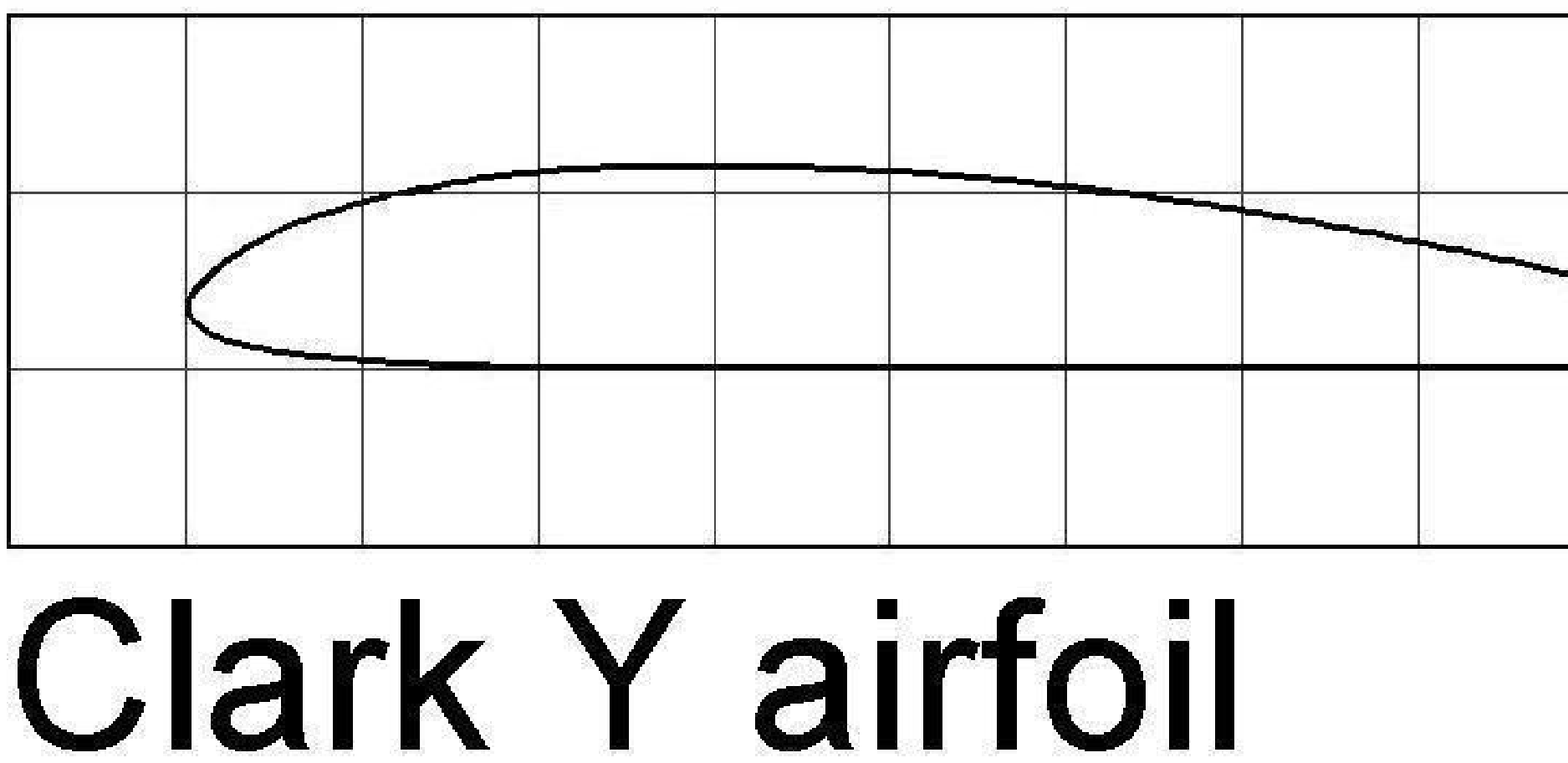}}%
\caption{Schematic of the curved wing and of the two airfoils analysed to validate the numerical code performances.
}
 \label{airfoils_sections}
 \end{figure*}

In this paper, we provide a preliminary set of data concerning the aerodynamics of an arc shaped, rigid, non twisted and non tethered wing which models an actual kite wing.
%; the data were obtained from numerical simulations of the Reynolds averaged turbulent flow.
% We have chosen the aspect 
The aspect ratio of the kite wing and the Reynolds number of the air flow, which gives a measure of the ratio of inertial forces to viscous forces, we have simulated are typical of current traction kite applications (aspect ratio $AR = 3.2$, Reynolds number based on the chord length $Re = 3 \times 10^6$, see e.g. \cite{Canale}). These data can be used to improve the design  of purpose-built kites for energy extraction and related flight control strategies. They can also represent a benchmark for comparison with  experimentally collected data and with any future unsteady flow simulations carried out by means of the large eddy simulation method.  The present simulations rely on the Reynolds Averaged Navier-Stokes model (RANS) and were carried out by using the  STAR - CCM+ Computational Fluid Dynamics code set up by CD-Adapco \cite{CD-Adapco}. This code solves the Navier-Stokes equations for an incompressible fluid using a finite volumes discretization. In order to verify the correct setting and flow features of the STAR - CCM+ code,  a comparison between the numerical and the laboratory characteristics of two archetype two-dimensional  airfoils  -- the NACA0012 and ClarkY profiles -- is also shown.

One of the aims of the paper is to obtain a comparison between the aerodynamics of a flat and a curved twin-set of rigid non-twisted wings with the same aspect ratio and the same Reynolds number, which is a topic that has not been discussed frequently in literature. The comparison is carried out by considering the boundary layer to be turbulent throughout. This choice slightly penalizes the prediction of the aerodynamic drag and can be considered a sort of systematic error we introduce {\it a priori} into the analysis to avoid the insertion of a non rational parametrization of the three dimensional transition on the kite, caused by the lack of reliable information about the three dimensional transition over curved three-dimensional non axial-symmetrical surfaces. 

%The results are produced using the standard computational techniques implemented in the widely used engineering code STAR-CCM+. \textit{We didn't try to optimize any of the computational aspects as, for instance, the structure of mesh grid or the turbulence model. TOGLIERE?}

In the context of kite dynamics, the present study, even though carried out by numerically simulating the three-dimensional turbulent viscous flow past a arc-shaped curved wing, should be considered of a preliminary nature. We did not address stability or optimization aspects, which
%One of our aims was in fact to offer an example of simple computation that could be easily reproduced by kite designers. 
could be considered in future works where the present physically comprehensive numerical simulations could be joined to stability and/or optimization techniques. It should be noted that, in this contest, a few papers have instead been published on simplified aerodynamic models of the complete system - i.e. kites and tethers. For example,  stability in a simplified simulation where the kite is a flat two-dimensional wing was considered by Alexander and Stevenson (2001) \cite{Alexander} , the dynamics of circular trajectories of a rigid flat kite was studied by Stevenson and Alexander (2006) \cite{Stevenson} , the optimization of the twist spanwise distribution - in the limit of a high wing aspect ratio - was addressed by Jackson (2005) \cite{Jakson} using the inviscid lifting line theory.

The paper is organized as follows. Next section is dedicated to the numerical simulation methodology. The third section presents the comparison between the aerodynamics properties of two wing sections, the Clark Y and the NACA0012 profiles, obtained from our simulations and those obtained from laboratory measurements. A fourth section is dedicated to the aerodynamics of the kite wing compared to the equivalent flat wing. We give also information about the mean pressure and kinetic energy fields along many kite sections and in the wake. The concluding remarks are in the last section.

%3
\begin{figure*}
 \centering
% \subfigure[Volume mesh along plane xy]
   {\includegraphics[width=\columnwidth]{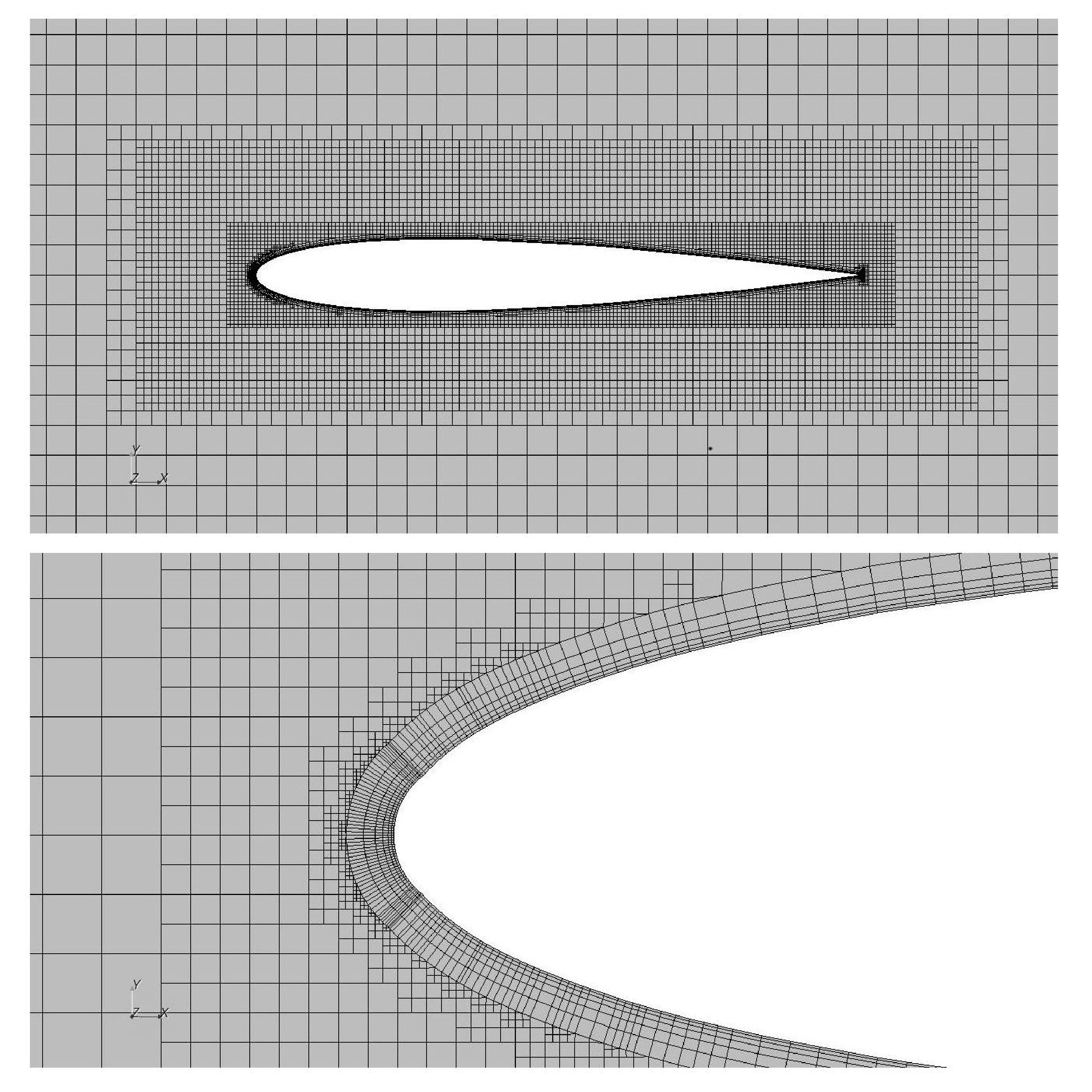}}
%  \quad%\hspace{4mm}
% \subfigure[Volume mesh along plane yz]
   {\includegraphics[width=\columnwidth]{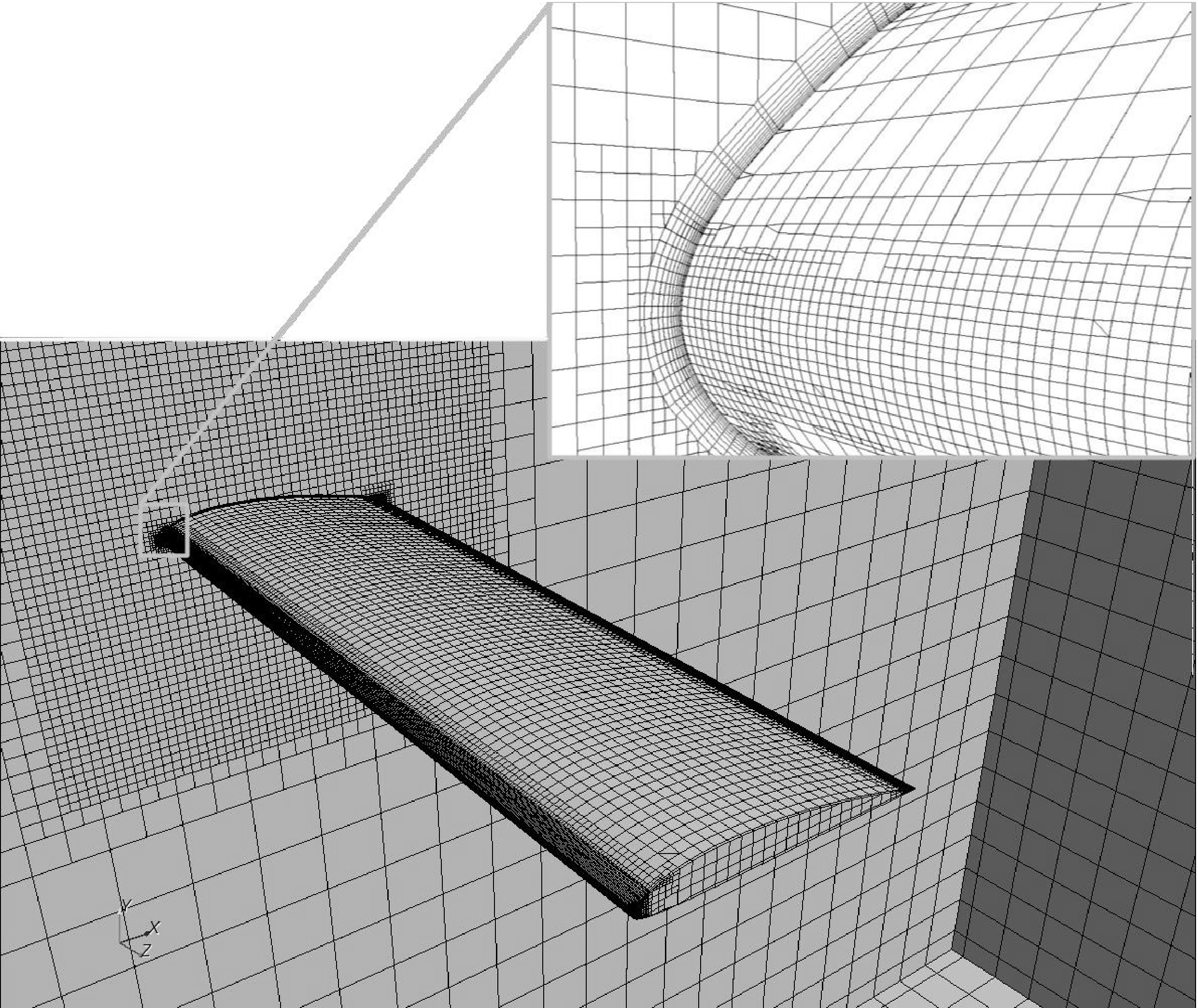}}%
 \caption{On the left: Naca0012 airfoil mesh ($Re = 3 \times 10^6$). On the right: the half flat wing volume mesh (Clark Y airfoil, $Re = 3 \times 10^6$, AR=6)}
 \label{volumemesh}
 \end{figure*}
%
%

%%%%%%%%%%%%%%%%%%%%%%%%%%%%%%%%%%%%%%%%%%%%%%%%%%%%%%%%%%%%%%%%%%%%%%%%
 \vspace{2mm}
\section{Numerical method}
The STAR-CCM+ industrial CFD code has been used to carry out the simulations. This code solves the Reynolds Averaged Navier-Stokes equations for an incompressible fluid using an unstructured, collocated finite-volume technique \cite{Ferziger}. The convection contribution to the velocity increment is predicted by an upwind scheme while a centered spatial discretization of the convection is introduced as a deferred correction (implicit pressure-correction method, SIMPLE \cite{Caretto} and SIMPLEC \cite{Doormal} algorithms). The Crank--Nicholson scheme is used for diffusion. The global scheme is thus second-order in space for steady state flows and, formally, first-order in time dependent flows.
%, but the truncation error limited to the velocity increment  is small.
This integration scheme is very stable, which is a necessary condition for a commercial code.

The problem symmetry, see figures \ref{kite_scheme}-\ref{airfoils_sections}, allows us to consider the infinite half-space by the side of the plane of symmetry of the wings as the computational domain. The domain boundaries are located three chord lengths upstream and six chord lengths downstream from the leading edge. The upper and lower boundaries are placed at five chord lengths each from the leading edge. The lateral boundary is located at three chord lengths from the wing tip.

Velocity and pressure are imposed on the domain inlet and uniformity conditions are imposed on the lateral boundaries and on the outlet. This includes the symmetry condition on the symmetry plane. The wing surface is treated like a rigid, non porous, wall where a no-slip condition applies.

The grid mesh is composed of an inner  layer surrounding the wing surface with a thickness that is suitable to capture the boundary layer. The mesh is particularly refined around the leading and trailing edges of the wing, while it is coarser on the remaining wing part of the surface. The outer mesh is composed of tetrahedral elements that become coarser towards the external boundaries of the computational domain, see figure \ref{volumemesh}.
The optimal cell density has been estimated by running several two-dimensional cases with a progressively increasing number of cells until a good agreement with laboratory data, obtained  from the existing literature, has been reached. The mesh has been extended along the wing span, while maintaining a similar density, until a final cell count close to $2.5 \times 10^6$ has been obtained, see figure \ref{volumemesh}.

The simulations were carried out using a commonly employed eddy-viscosity turbulence model, the  turbulent viscosity transport equation model by Spalart and Allmaras \cite{Spalart}. This model was built using heuristics and dimensional analysis arguments. The transport equation  is local, which means  that the equation at one point does not depend on the solution at the neighbouring points, and it includes a non-viscous reduction term that depends on the distance from the wall. This property makes the model compatible with grids of any structure. A laminar-turbulence transition was not imposed, the boundary layer was considered turbulent throughout. This choice was motivated, on the one hand, by the desire to  avoid the inclusion of  uncertain parameters linked to the as yet unknown transition dynamics on curved three-dimensional non axial-symmetrical surfaces, and, on the other, because of the awareness that the associated overestimation of the drag coefficients implies results on the safer side.

\begin{figure}
 \centering
{\includegraphics[width=\columnwidth]{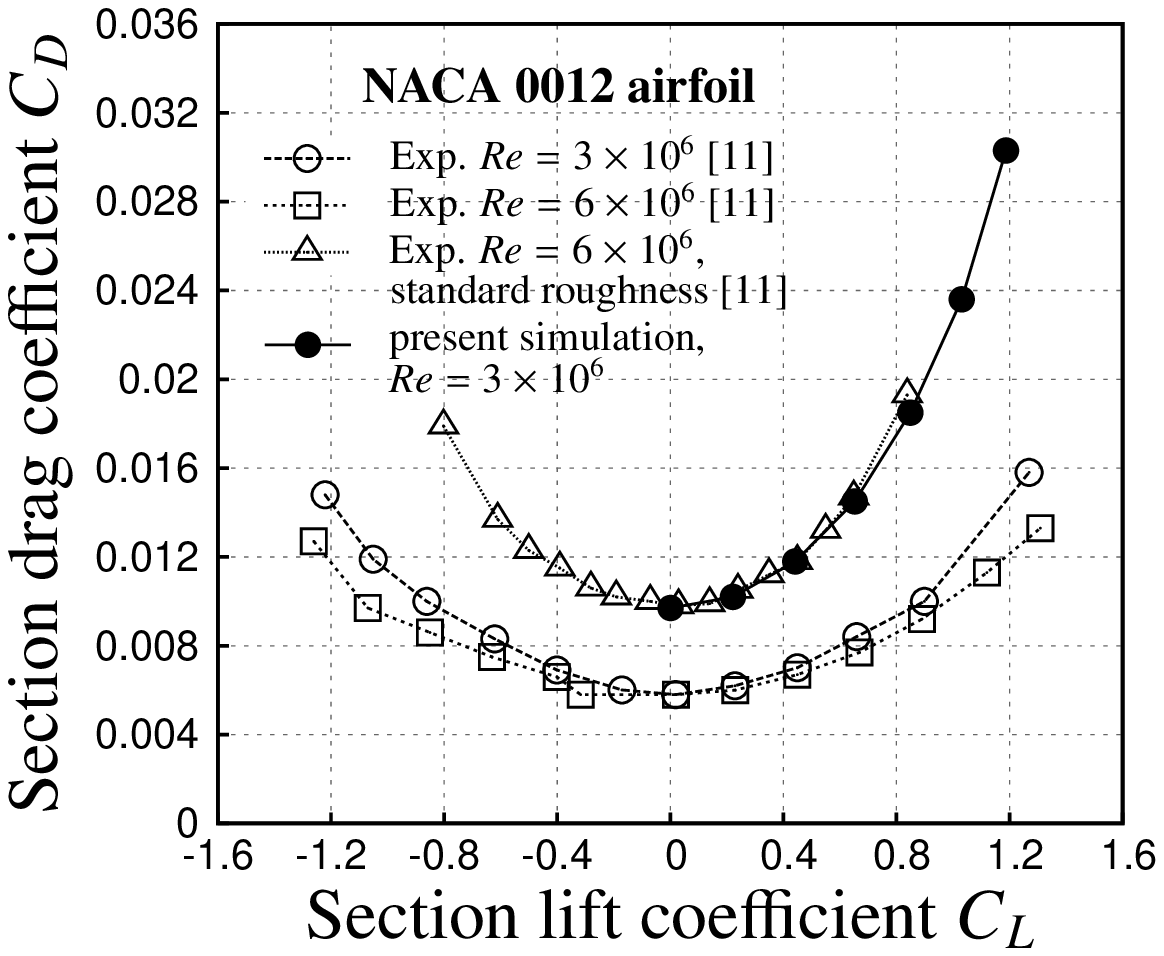}}\\
{\includegraphics[width=\columnwidth]{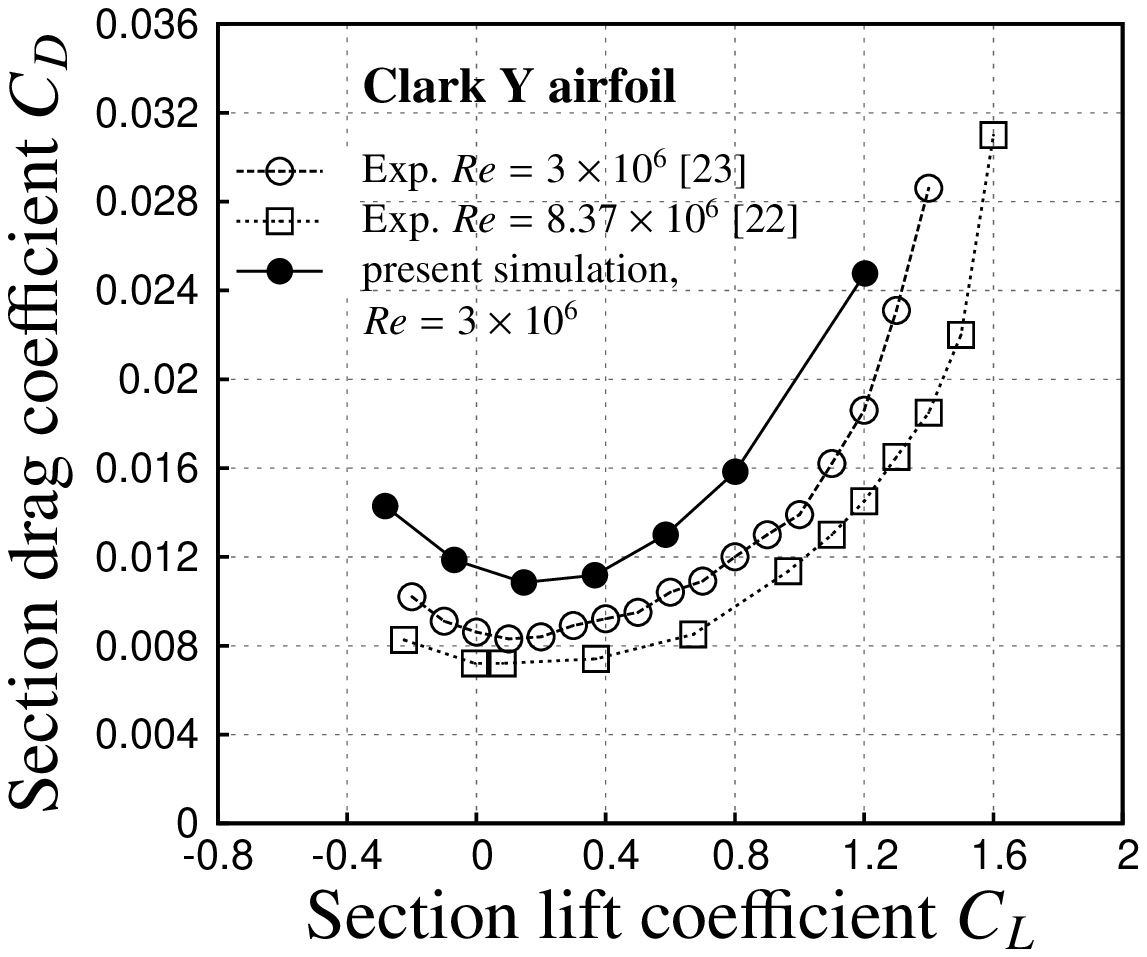}}
% {\includegraphics[width=0.50\textwidth]{CD_CL_NACA0012Airfoil.eps}}%
%  %\quad%\hspace{4mm}
% \vspace{-2mm}
% {\includegraphics[width=0.50\textwidth]{CD_CL_ClarkYAirfoil.eps}}
% \vspace{-20mm}
% \caption{Comparison between the polar curves of the NACA0012 and Clark Y profiles, obtained from the present numerical simulations and from laboratory experiments in literature.}
\caption{Comparison between the polar curves of the NACA0012 and Clark Y profiles obtained from the present numerical simulations (STAR-CCM+ code with the Spalart-Allmaras turbulence model) and from laboratory measurements in literature: Abbott and von Doenhoff \cite{Abbott}, Silverstein \cite{R502} and Jacobs and Abbott \cite{TR669}.}
 \label{2Dpolars}
 \end{figure}

%5
\begin{figure}
 \centering
% %  \subfigure[Cl-Alfa]
%    {\includegraphics[width=0.48\textwidth]{CL_Alfa_Comparison.ps}}%
%  \quad%\hspace{4mm}
% %  \subfigure[Polar curve: comparison of aerodynamic coefficients between kite and rectangular wings]
%    {\includegraphics[width=0.48\textwidth]{CD_CL_Comparison.ps}}%
%  \subfigure[Cl-Alfa]
   {\includegraphics[width=\columnwidth]{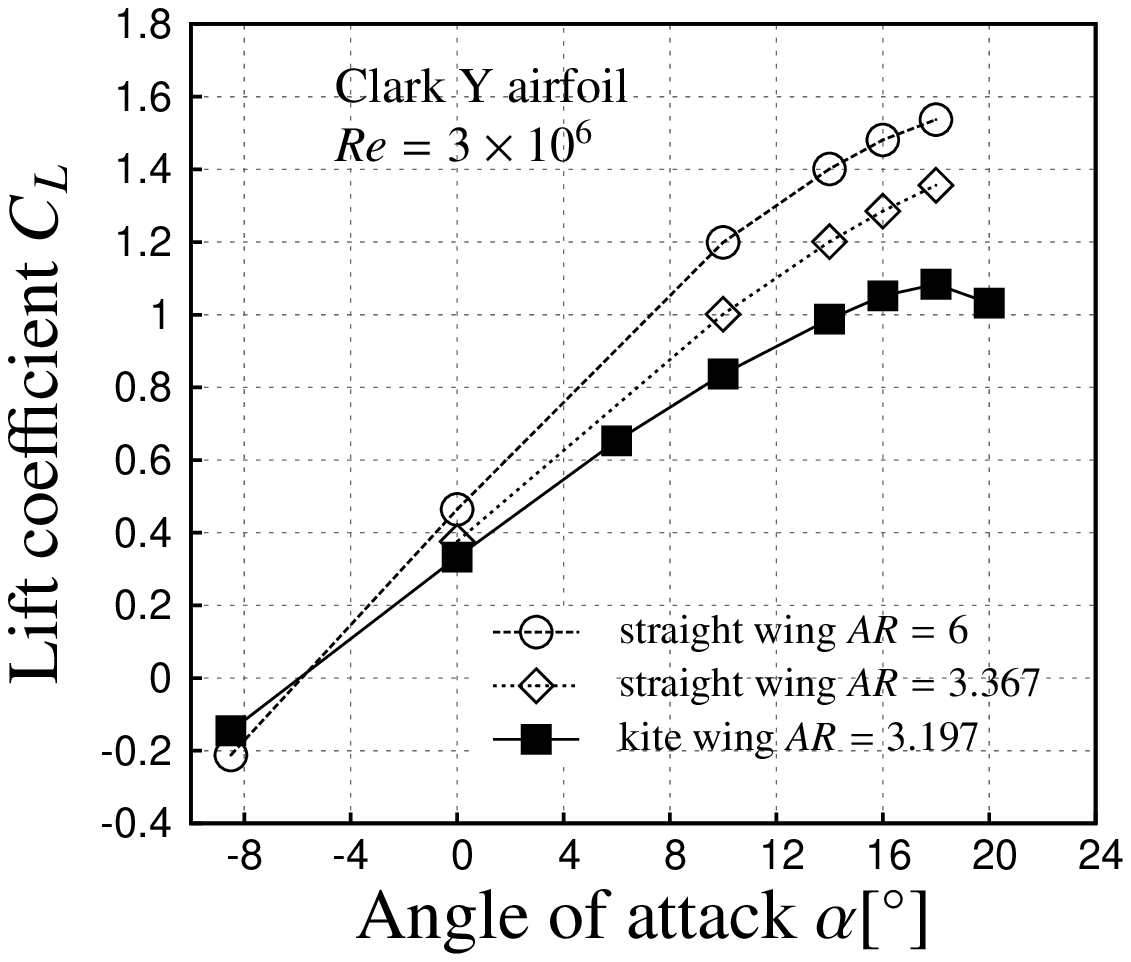}}\\ %
%  \quad%\hspace{4mm}
%  \subfigure[Polar curve: comparison of aerodynamic coefficients between kite and rectangular wings]
   {\includegraphics[width=\columnwidth]{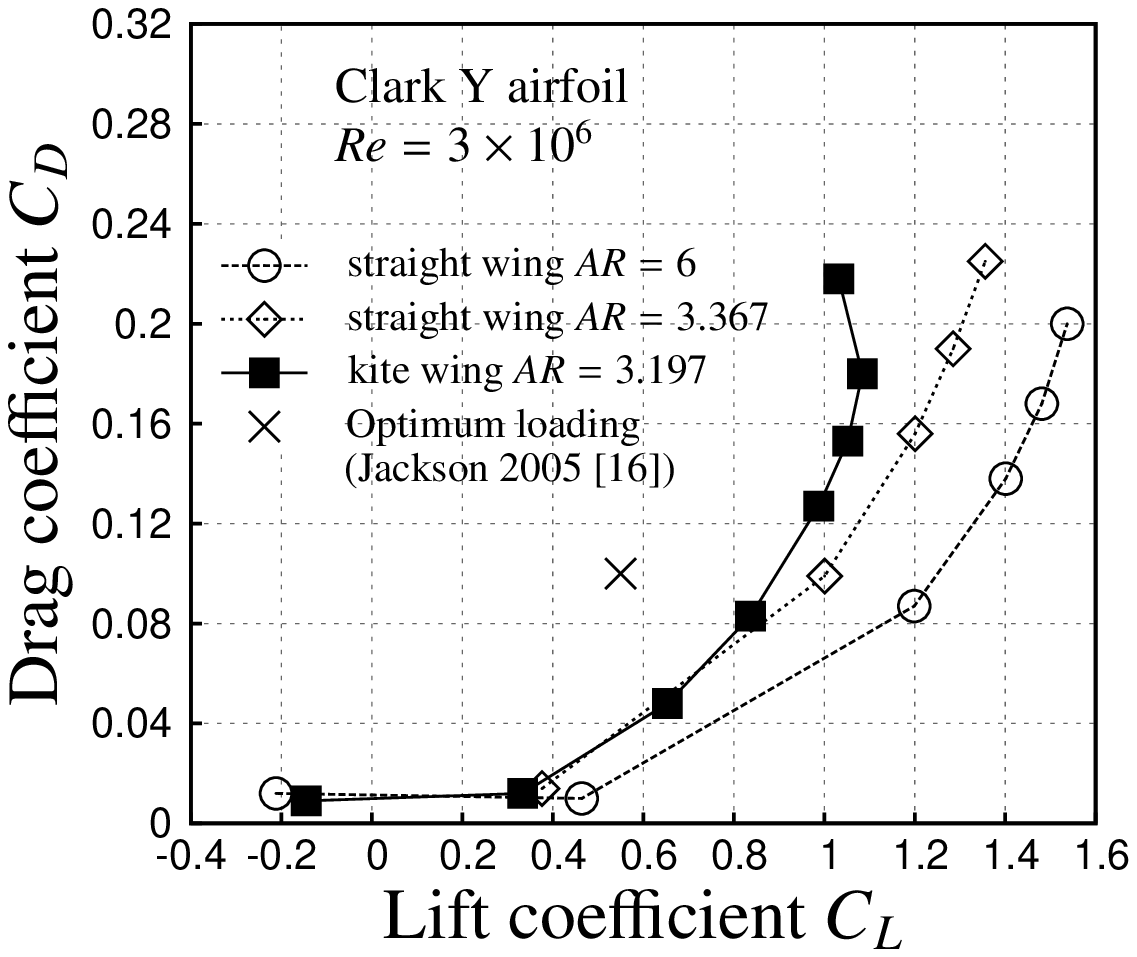}}%

%  \caption{Aerodynamics characteristics of the curved wing contrasted with the equivalent flat wing. $Re=3 \cdot 10^6$ based on the chord length, $AR = 3.2$.}
\caption{Aerodynamics characteristics of the curved wing contrasted with the equivalent flat wing. The Reynolds number, based on the chord length, is equal to $3 \times 10^6$. All simulations have been carried out with the STAR-CCM+ code using the Spalart-Allmaras turbulence model. The optimum loading for a tension kite from \cite{Jakson} is shown as a reference.} %, $AR = 3.2$.}
% {Jakson}
 \label{kite_polars}
 \end{figure}

% \vspace{2mm}

{\section{Validation of the numerical results.\\ The NACA0012  and Clark Y airfoil test cases.}}

The numerical results produced using the STAR-CCM+ code have been validated through a comparison with  results produced in the laboratory for two important two-dimensional test cases, the incompressible flows past the NACA0012 and the Clark Y profiles, at various angles of attack.
In order to obtain  a set of  numerical results that are consistent with laboratory results in a range of different flow conditions  it is  necessary  to achieve sufficient confidence in the use of the code parameter setting. This setting is of utmost importance as it specifies the physical model of the problem under study.

For this reason, we analysed the flow around two wing sections for which a large amount of laboratory literature is available. The NACA 0012 is a symmetrical airfoil with a maximum thickness of 12\%\ of the chord. This is probably the most extensively studied wing section. The Clark Y is an asymmetrical profile with a flat bottom. This profile is the section proposed for the kite wing. An extended laboratory database was also available for this section.

The  flow past the two wing sections has a Reynolds number  $Re$ of $3 \times 10^6$. The angle of incidence  $\alpha$ is varied from $0^{\rm o}$ to $20^{\rm o}$. The $Re$ is based on the chord length, $c=1$ m, and on the freestream velocity, $U_{\infty}=\;43,86$ m/s. The air density is set to $\rho=1,225$ kg/m$^3$, the pressure to $p_{\infty}=101325$ Pa, the dynamic viscosity to $\mu=1,79\times 10^{-5}$ poise and the temperature to $T_{\infty}=288$ K. The boundary layer is considered turbulent throughout, a condition which is physically close to the flow configuration experimented for sections with a non zero surface roughness.

The domain extends three chords in front of the profile, six chords downstream from the leading edge and ten chords in the transversal direction. The mesh consist of 30105 cells. % for the simulation over the NACA0012  and of 64041 cells for the the Clark Y section. 
The mesh is refined through prism layers in the proximity of the airfoil. Approaching the stall condition, the number of iterations has been increased, from about 1000 to about 1800, to grant solution convergence. This criterion was also applied to  the three-dimensional simulations, where the number of iterations necessary to converge resulted to be of the same order of magnitude. 

The results concerning the polar curve for the NACA0012 airfoil were compared with the experimental measures reported by Abbott \& Doenhoff \cite{Abbott}. Figure 4 shows the numerical prediction of the lift and drag coefficients of the airfoils compared with the laboratory data. These coefficients are defined as $C_l=L/(\rho U_{\infty}^2 c/2)$ and $C_d=D/(\rho U_{\infty}^2 c/2)$ where $L$ and $D$ are the lift and drag forces per unit length and $c$ is the chord length.
As expected, since the boundary layer was considered turbulent along the entire profile, the agreement is excellent  with the data relevant to the airfoil with standard roughness. The agreement becomes more qualitative with regards the data relevant to smooth profiles. As previously explained, in this last case, the numerical prediction of the drag coefficient is biased. It can in fact be observed that the setting for the boundary layer  of a turbulence ubiquity condition induces an over-estimation of the drag of about a 40\%. However, it is important to note that this bias does not spoil the parallelism between  the laboratory and numerical polar curves. 

For the Clark Y airfoil, the numerical prediction of the aerodynamic characteristics was compared with the  wind tunnel measurement  carried out in the Langley variable-density tunnel ($Re = 8.37 \times 10^6$,  \cite{TR669}) and in the full-scale tunnel ($Re = 3 \times 10^6$, \cite{R502}). It should be noted that, for the Clark Y profile, data for rough surfaces are not present in literature. Even though old, we decided to consider these experimental data because they were not produced in a low-turbulence tunnel. As a consequence, in principle, these data can offer closer {\it a priori} estimate of the flow configuration over a  profile where the boundary layer is turbulent throughout. Fig ure\ \ref{2Dpolars} shows a comparison between the numerical and laboratory experiments. As above, it can be seen that the numerical polar is parallel to the laboratory one. Again in this case, the numerical polar over-predicts the drag coefficient by about a 40$\div$50\%.

In conclusion, we can observe a good trend of the polar curves, which are parallel to the experimental ones. The drag is over-estimated, as it should be, since we decided not to enlarge this preliminary study of the kite wing aerodynamics with other parameters to fix the transition line on the wings. The drag over-estimation should therefore be considered a consistent result.

% \section{Aerodynamics of the curved kite wing compared to the equivalent flat wing.  Related characteristics of the flow field}

% \section{Aerodynamics of the curved kite wing: comparison with the equivalent flat wing and related characteristics of the flow field}

%6
\begin{figure*}
 \centering
    {\includegraphics[width=\columnwidth]{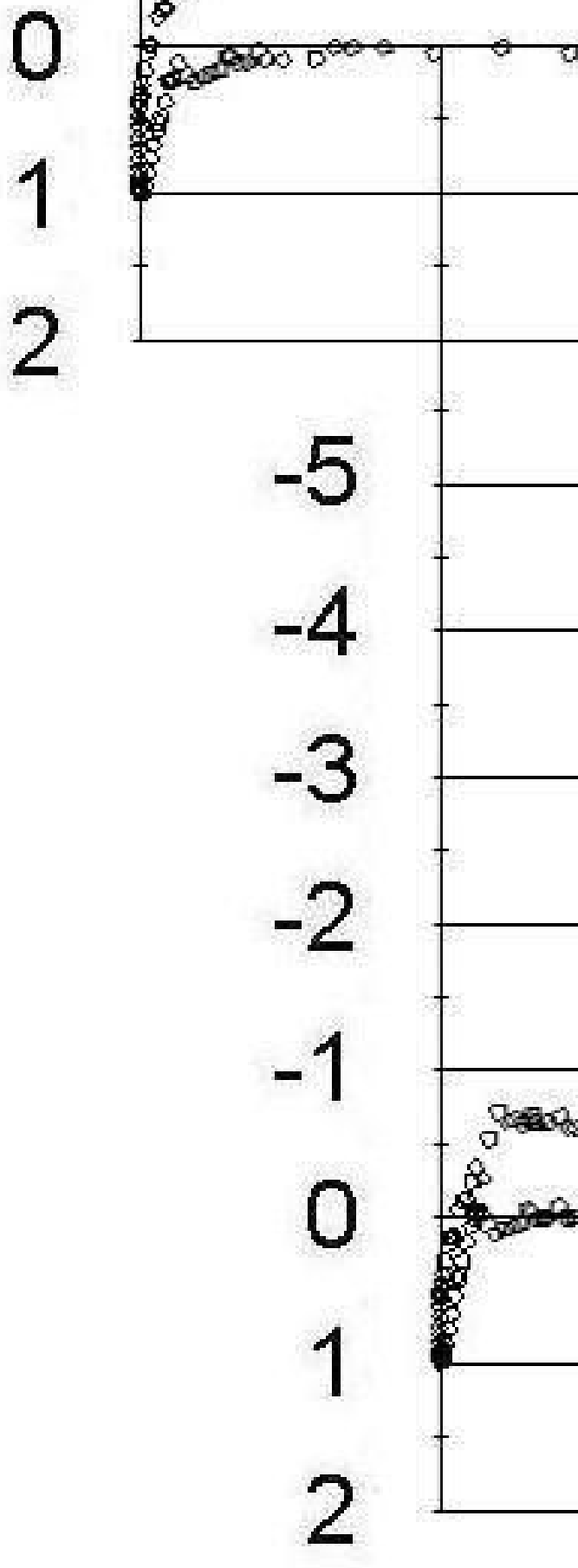}}%
\quad%
    {\includegraphics[width=\columnwidth]{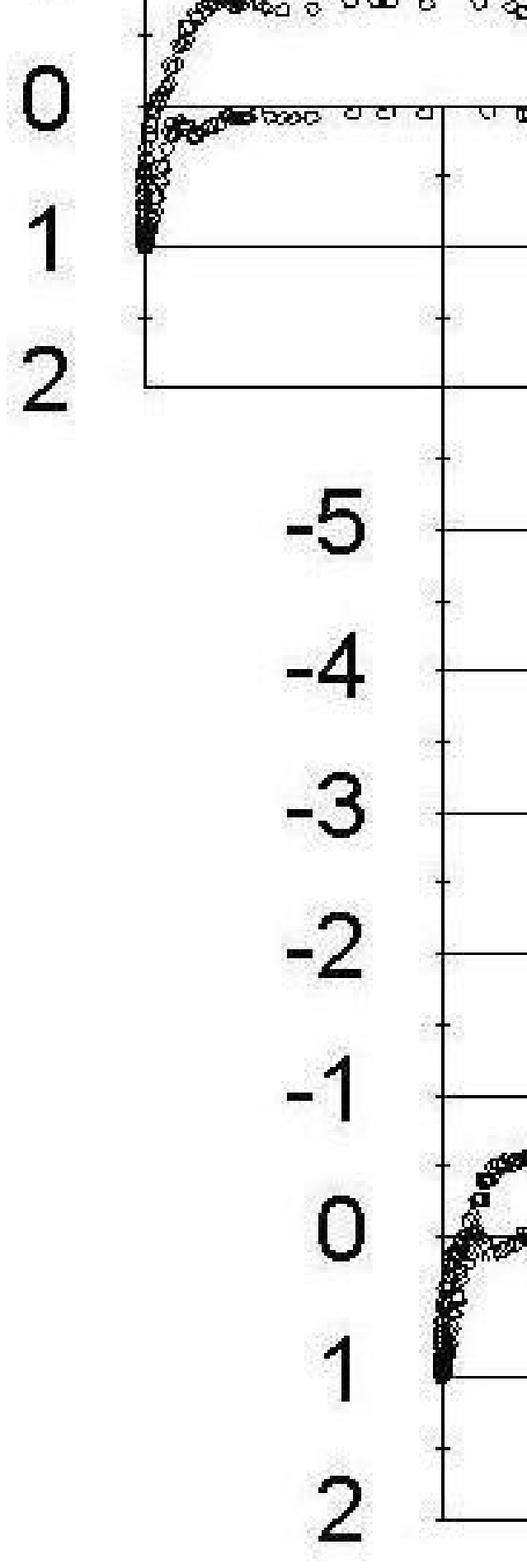}}%
\caption{Pressure coefficient distributions at different $z=const$ sections along the flat wing (part a). Pressure coefficient distributions at different $\theta=const$ sections along the curved wing (part b). The angle of incidence is $6^{\rm o}$; $Re=3 \times 10^6$ based on the chord length, $AR = 3.2$.}
 \label{Cp_6}
 \end{figure*}

\begin{figure*}
 \centering
{\includegraphics[width=\columnwidth]{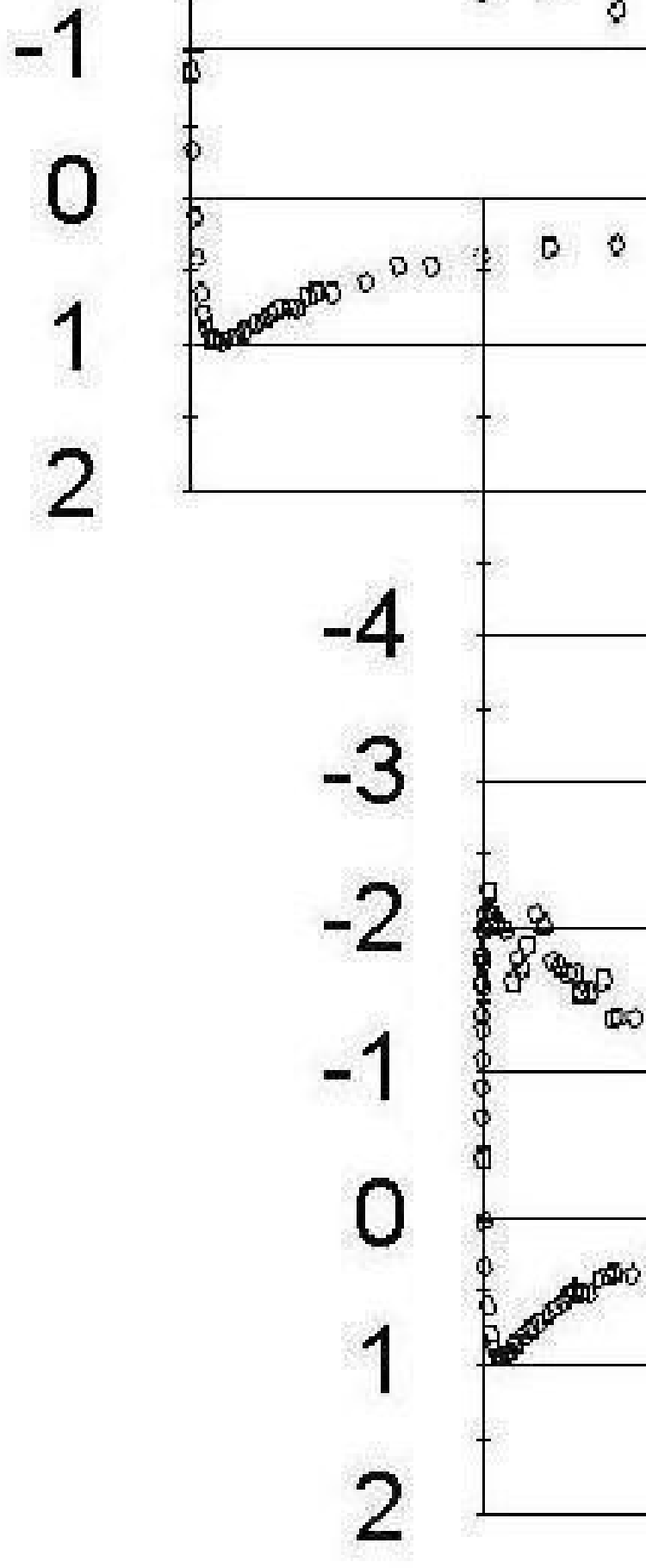}}%
\quad%
    {\includegraphics[width=\columnwidth]{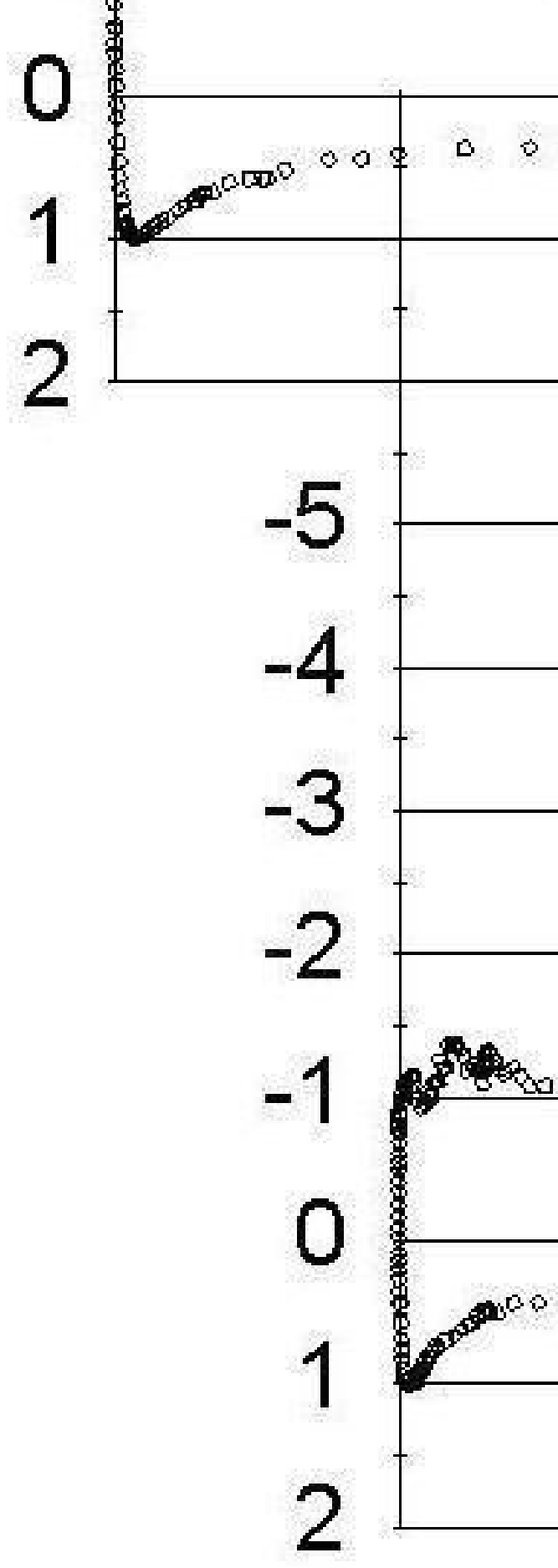}}%
%{\includegraphics[width=0.80\textwidth]{pressure-distrib-straightw.eps}}%
\caption{
Pressure coefficient distributions at different $z=const$ sections along the flat wing (part a). Pressure coefficient distributions at different $\theta=const$ sections along the curved wing (part b).  The angle of incidence is $18^{\rm o}$; $Re = 3 \times 10^6$ based on the chord length, $AR = 3.2$}
\label{Cp_18}
\end{figure*}

\vspace{2mm}
\section{Aerodynamics of the curved kite wing and modification of the flow past the wing. Comparison with the equivalent flat wing}

We present two main sets of results. The first is associated to the polar curves $C_L=C_L(\alpha)$,  $C_D=C_D(\alpha)$ and the pressure distribution on the upper and lower surfaces of both the flat and curved wings.
% ($C_L=L/\frac{1}{2}\rho U_{\infty}^2 c$ and $C_D=D/\frac{1}{2}\rho U_{\infty}^2 S$ are the lift and drag coefficients of the wing, $S$ is the reference surface formed by the set of wing chords)
These data specify the aerodynamic characteristics of the two kinds of wings. We adopted the Clark Y airfoil as wing section for both the curved kite wing and the straight wing. The aerodynamic lift and drag coefficients are formed using the reference  force $1/2 \rho S U_{\infty}^2$, where $S$ is half the net surface (or the surface formed by the set of wing chords). The Reynolds number, based on the chord length and the free stream air velocity, is fixed and is equal to $3 \times 10^6$. The polar curves  are shown in figure \ref{kite_polars}, while the pressure distribution in part (a) - the flat wing, and part (b) - the curved wing - are shown in figures \ref{Cp_6} and \ref{Cp_18}. 

As explained before, see the introduction and following sections, the drag coefficients are biased (overestimated by about 40-50 \%)) due to the turbulence ubiquity flow conditions that were adopted in the boundary layer in the absence of a reliable criterion to approximate the three-dimensional separation transition on the curved surfaces of the kite wing. 
The polars were obtained by varying the angle of attack in the $[-8^{\rm o}$ to $20^{\rm o}]$ range. Figure \ref{kite_polars} shows the polar curves for two flat wings with aspect ratios equal to $3.37$ and $6$, and for the kite wing with an aspect ratio of $3.2$. It can be observed that the characteristics are very close in the angle of attack range $[-8^{\rm o}, 4^{\rm o}]$, to which a range of lift coefficients from $-0.2$ to $0.6$ corresponds. As expected, the slope of the lift coefficient curve deteriorates (decreases) a little moving from the flat wing with  $AR=6$ to the flat wing with $AR=3.37$, and from the latter to the curved wing with $AR=3.37$. Beyond an angle of attack of about $4^{\rm o}$, the lift coefficient curve for the curved wing starts to bend. Stall is reached at an angle of attack of $18^{\rm o}$, where $C_L = 1.1$ and $C_D = 0.18$.  It should be noted that the flat wings have not stalled yet at $18^{\rm o}$ of incidence.
An analogous behaviour is shown by the $C_D =  C_D(\alpha)$ polar curve, which deteriorates by reducing the aspect ratio and by switching  from the flat to the curved configuration. This curve  contains the information on the  aerodynamic efficiency. For the flat  wing with $AR=6$, the maximum efficiency  is $40$. This value reduces to $26$ for both the flat and the curved wings ($AR \sim 3$).

The set of pressure distributions on the lower and upper wing surfaces also describes the lift distribution on the wing. The pressure is normalized in the form of pressure coefficients $C_p$ ($C_p = p - p_{\infty} / (1/2 \rho U^2_{\infty}$), where the suffix $\infty$ means the asymptotic upstream condition, $\rho$ is the density and $U_{\infty}$ the free stream speed). The two wing distributions are compared in figures \ref{Cp_6} and \ref{Cp_18} at $\alpha = 6^{\rm o}$ and 
$\alpha = 18^{\circ}$, respectively. The distributions agree with the behaviour observed in the laboratory on a two-dimensional wing with a Clark Y section \cite{Wenzinger}. In particular, with the exclusion of the wing tip region, a slightly negative pressure value is observed at the trailing edge, which is a typical feature of the Clark Y profile. 
The  position along the span on the curved and on the flat wing is measured in terms of the angle $\theta$ and of the value of the $z$ coordinate, respectively in figures \ref{Cp_6} and \ref{Cp_18}, see figures \ref{kite_scheme} and \ref{airfoils_sections}. The correspondence is made in such a way that the corresponding sections have the same value as the curvilinear (for the curved wing) or rectilinear (for the flat wing) coordinates running along the wing span. It can be noted that, at the lower angle of attack of $6^{\circ}$, see Figure \ref{Cp_6}, the flat wing and the curved wing show similar pressure distributions at the corresponding sections. The distributions are almost equal on the lower surface. On the higher surfaces, but only in the region near  the leading edge,  a $10-15\%$ of difference is noted  in the central part of the wing, a value which increases to 40-50\%\ in the wing tip region. The same trend can be observed at the angle of attack of $18^{\circ}$, see figure \ref{Cp_18}. However, the difference in the pressure values at the leading edge now rises to values close to 15\%\ in the central part of the wing and close to 100\%\ in the wing tip region, see also the flow visualization in figures \ref{kite_streamlines_alfa6} and \ref{kite_streamlines_alfa18}.

%8
\begin{figure*}
 \centering
    {\includegraphics[width=0.65\textwidth]{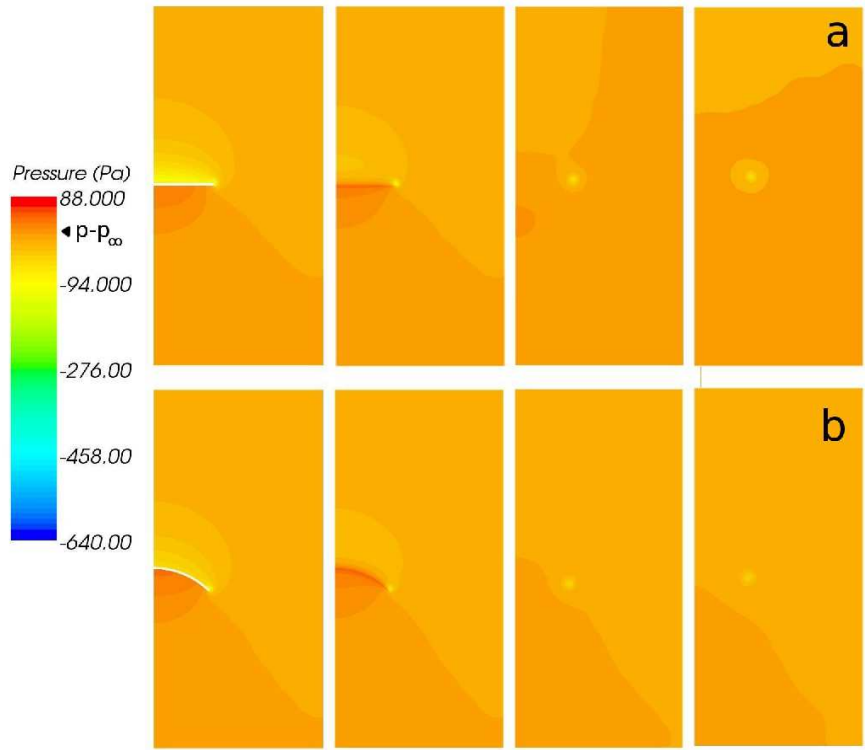}}%
% {\includegraphics[width=0.80\textwidth]{averaged-kinetic-energy-distrib_kite.eps}}%
\caption{Pressure levels in flow sections ($y, z$) across the wing and the wake at 3/4, 1, 2 and 3  chord lengths. Part a) flat wing. Part b) curved wing. The angle of incidence is $6^{\rm o}$; $Re=3 \times 10^6$ based on the chord length, $AR = 3.2$, $p_\infty = 101325$ Pa.}
 \label{PA6}
 \end{figure*}

%9
\begin{figure*}
 \centering
    {\includegraphics[width=0.65\textwidth]{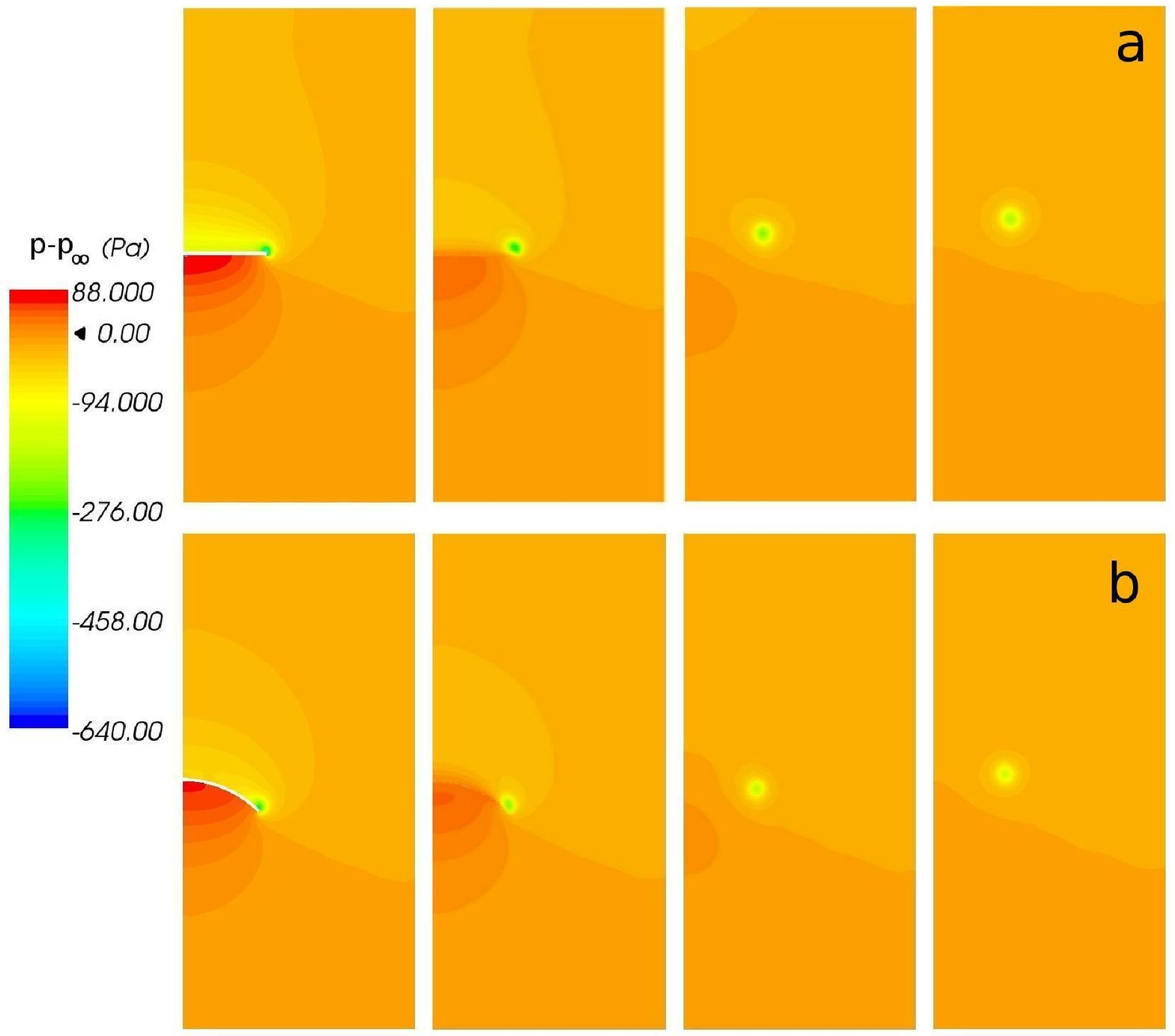}}%
% {\includegraphics[width=0.80\textwidth]{averaged-kinetic-energy-distrib_kite.eps}}%
\caption{Pressure levels in flow sections ($y, z$) across the wing and the wake at 3/4, 1, 2 and 3 chord lengths. Part a) flat wing. Part b) curved wing. The angle of incidence is $18^{\rm o}$; $Re=3 \times 10^6$ based on the chord length, $AR = 3.2$, $p_\infty = 101325$ Pa.}
 \label{PA18}
 \end{figure*}

%10
\begin{figure*}
 \centering
    {\includegraphics[width=0.65\textwidth]{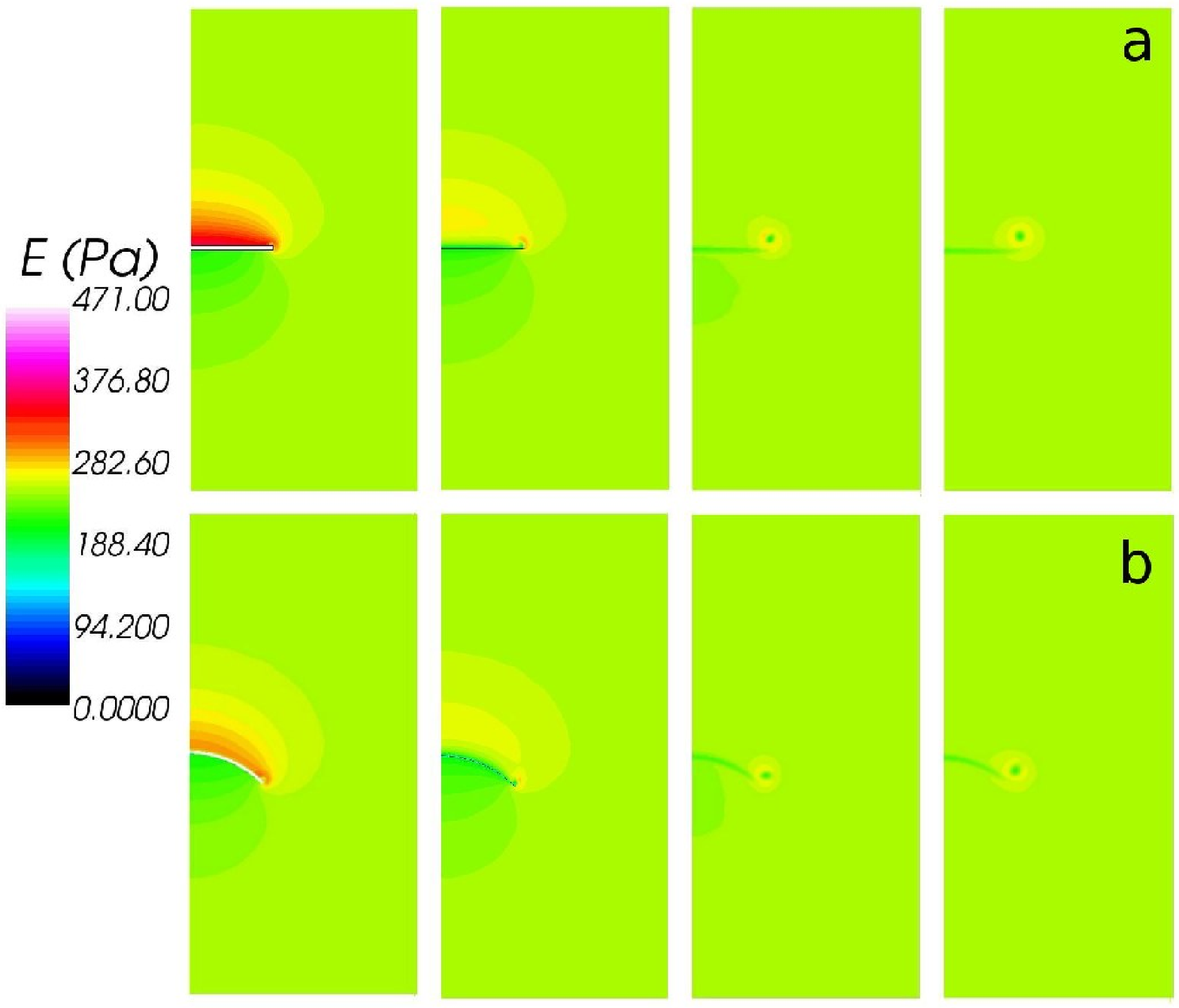}}%
% {\includegraphics[width=0.80\textwidth]{averaged-kinetic-energy-distrib_kite.eps}}%
\caption{Averaged kinetic energy levels in flow sections ($y, z$) across the wing and the wake at 3/4, 1, 2 and 3 chord lengths downstream from the leading edge. Part a) flat wing. Part b) curved wing. The angle of incidence is $6^{\rm o}$; $Re=3 \times 10^6$ based on the chord length, $AR = 3.2$, $E_\infty=247.45$ Pa.}
 \label{EA6}
 \end{figure*}

%11
\begin{figure*}
 \centering
    {\includegraphics[width=0.65\textwidth]{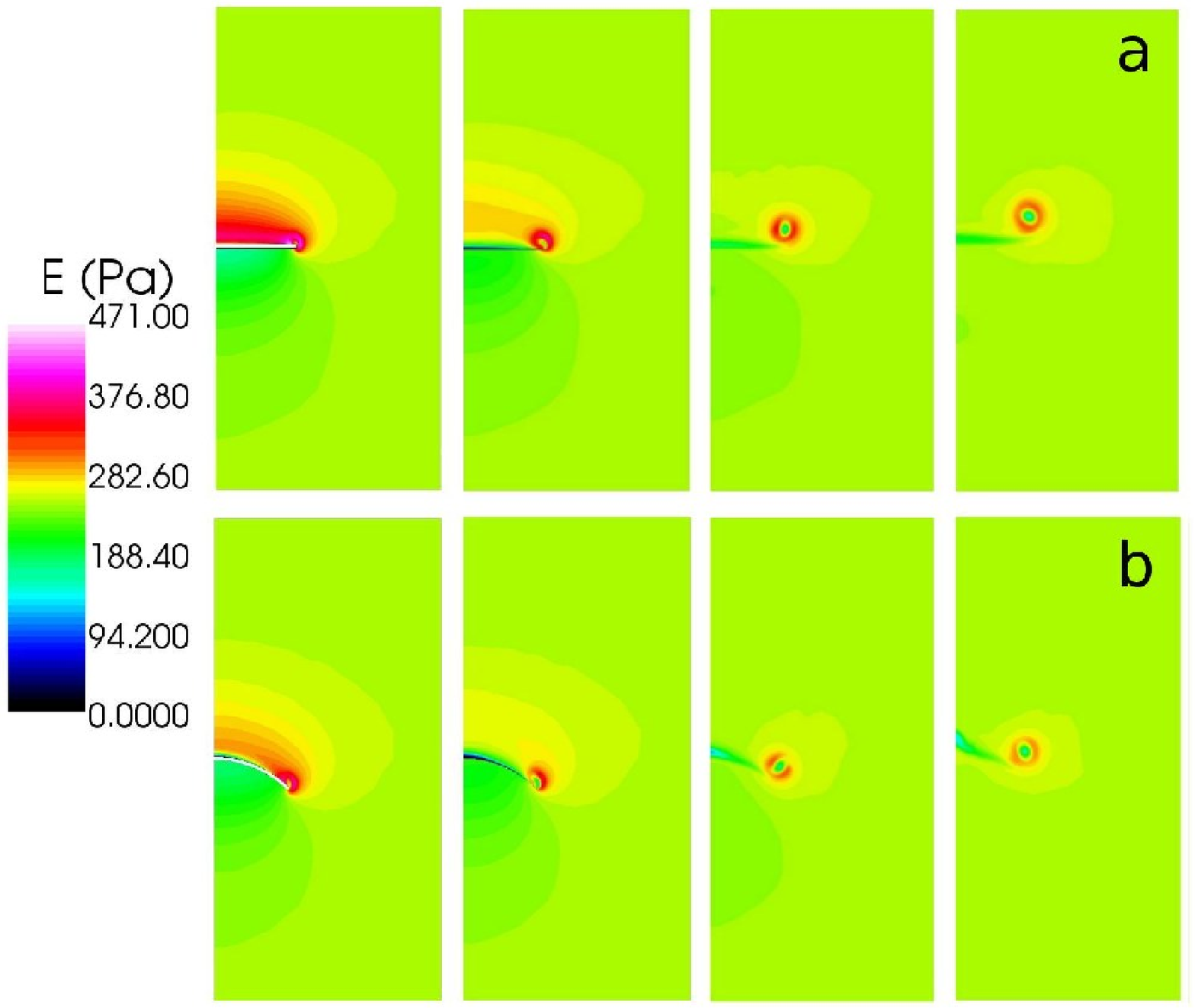}}%
% {\includegraphics[width=0.80\textwidth]{averaged-kinetic-energy-distrib_kite.eps}}%
\caption{Averaged kinetic energy levels in flow sections ($y, z$) across the wing and the wake at 3/4, 1, 2 and 3  chord lengths downstream from the leading edge. Part a) flat wing. Part b) curved wing. The angle of incidence is $18^{\rm o}$; $Re=3 \times 10^6$ based on the chord length, $AR = 3.2$; $E_\infty=247.45$ Pa.}
 \label{EA18}
 \end{figure*}

%12
\begin{figure*}
 \centering
    {\includegraphics[width=0.65\textwidth]{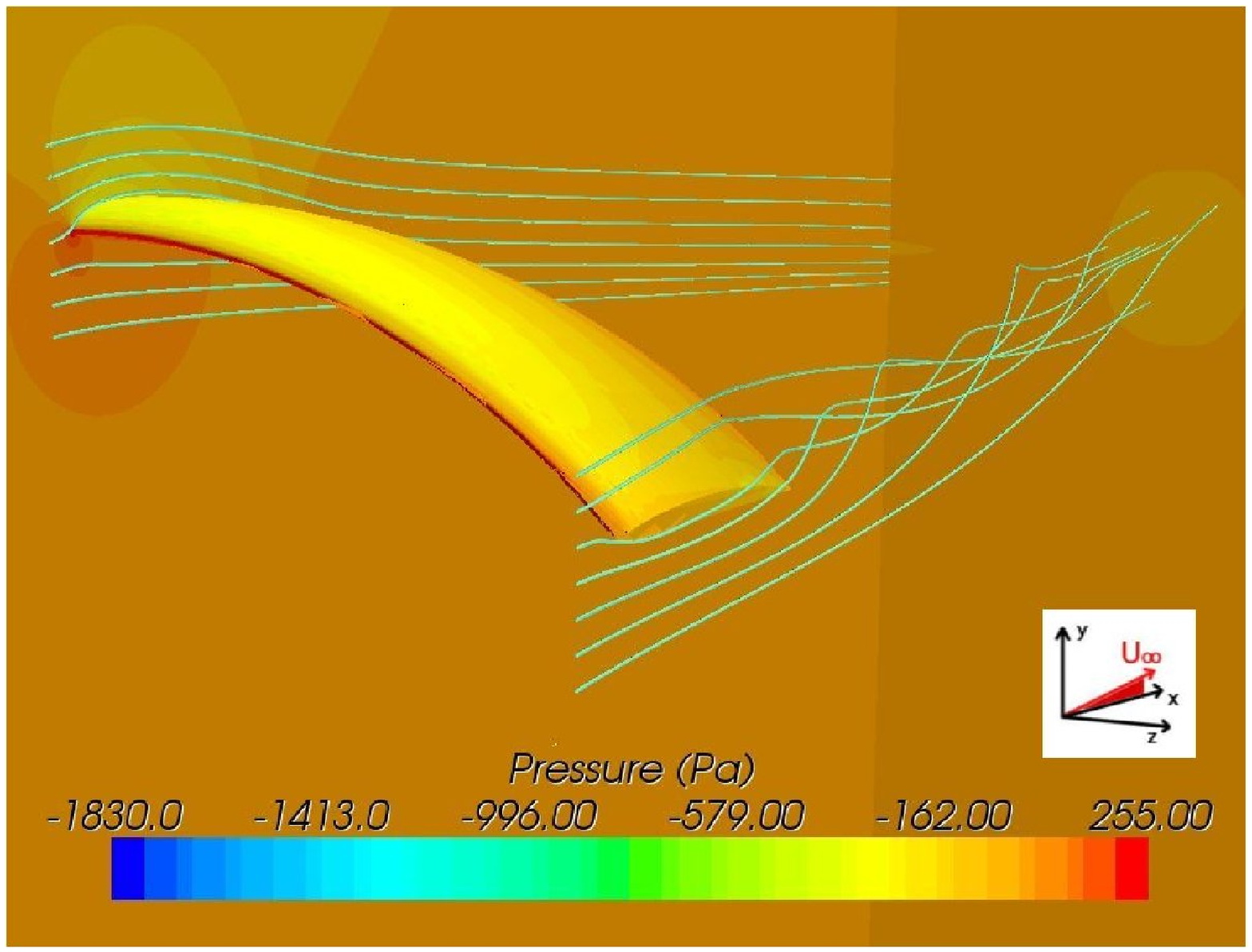}}%
\caption{Flow streamlines for the curved wing at an incidence of $6^{\rm o}$; the streamline visualization is associated to the pressure levels obtained on the wing surface, the symmetry plane and outlet boundary. $Re=3 \times 10^6$ based on the chord length, $AR = 3.2$. Note that the $x$ direction is the chord direction, see figure \ref{kite_scheme}, and it is parallel to the lateral domain boundaries.}
 \label{kite_streamlines_alfa6}
 \end{figure*}

%13
\begin{figure*}
 \centering
    {\includegraphics[width=0.65\textwidth]{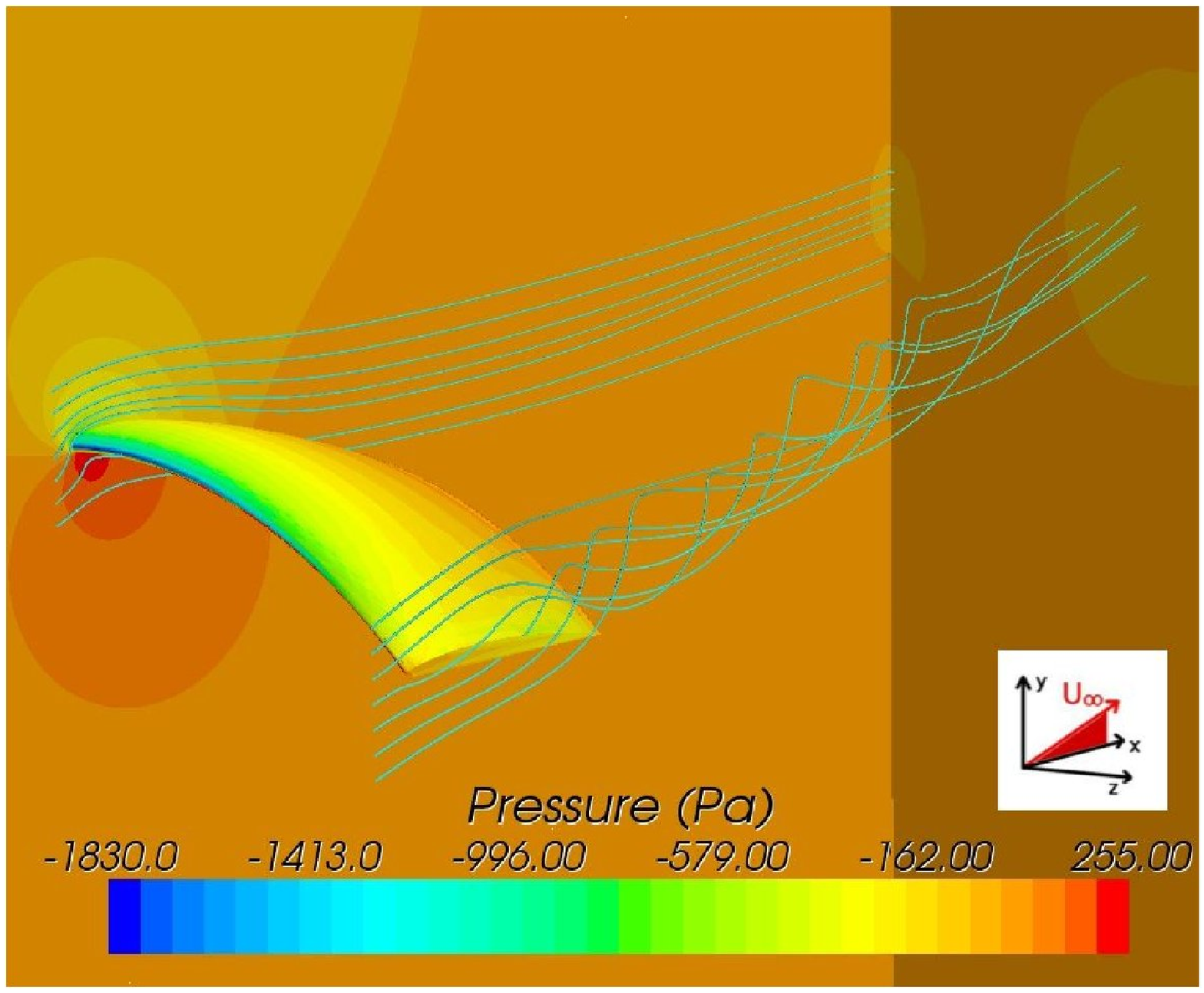}}%
\caption{Flow streamlines for the curved wing at an incidence of $18^{\rm o}$; the streamline visualization is associated to the pressure levels obtained on the wing surface, the symmetry plane and  outlet boundary. $Re=3 \times 10^6$ based on the chord length, $AR = 3.2$. Note that the $x$ direction is the chord direction, see figure \ref{kite_scheme}, and it is parallel to the lateral domain boundaries.}
 \label{kite_streamlines_alfa18}
 \end{figure*}

The second set of results concern information on the structure of the flow field, in particular, on the pressure and kinetic energy fields, see figures \ref{PA6} - \ref{kite_streamlines_alfa18}.
The pressure field visualization in figures \ref{PA6} and \ref{PA18} shows that the pressure distributions are qualitatively similar for the two kinds of wings. The pressure drop above the wing and inside the wing tip vortices is more intense at the higher angle of attack. At a three chord length behind the leading edge, for both angles of attack,  the pressure becomes almost uniform, with the exclusion of the traces of the tip vortices. A similar overall behaviour can also be observed  for the kinetic energy field, see figures \ref{EA6} - \ref{EA18}. Most of the averaged kinetic energy is concentrated above the wing  and just outside the wing tip vortex cores. Above the curved wing, for both angles of attack, a comparatively small kinetic energy, with regards to the flat  configuration, can be seen. It means that, in this section (3/4 of the chord), the flow on the curved wing has already separated, a fact that can be also deduced from the pressure distribution near the wing ends in figures \ref{Cp_6} and \ref{Cp_18}. 

Furthermore, if we consider the vorticity in the wake and, in particular, that in the tip vortices,  interesting observations can be made. For instance, in the 3/4 chord section, the vorticity produced by the flat wing is 1.3 times that produced by the curved wing, at $\alpha=6^{\rm o}$, a value that decreases to 1.075 at  $\alpha=18^{\rm o}$. However, if we compare the vorticity of the wing tips at $6^{\rm o}$ to that at $18^{\rm o}$, a ratio of 0.78 can be observed for the flat wing and of 0.64 for the curved wing. By moving downstream  the section at a 3 chord lengths the vorticity produced by the curved wing becomes slightly higher than that produced by the flat wing (flat/curved  yields ratios of 0.99 at $6^{\rm o}$ and 0.85 at $18^{\rm o}$).  This is in agreement with the fact the kinetic energy at 3 chords downstream is about 22\%\ of that on the wing for the flat configuration, and is about  28\%\ for the curved configuration. These observations mean that the curvature is less efficient in increasing the vorticity than the increase of angle of incidence, but it is capable, due to non-linear convective effects, of inducing slower spatial decay in the near wake.  

Another interesting point is that, by changing the angle of incidence, the convergence of the tip vortex axes remains almost constant ($2.1^{\rm o}$) for the flat wing, while it increases for the curved wing ($1.9^{\rm o}$ at $6^{\rm o}$ of angle of attack, $2.5^{\rm o}$ at $18^{\rm o}$ of angle of attack). This can also be seen also in figures \ref{kite_streamlines_alfa6} and \ref{kite_streamlines_alfa18} observing the visualization of the streamlines of the tip vortices. Thus, it can be concluded that  the curvature induces  more intense non linear effects (convection and stretching) on the vorticity and, as a consequence the vortices keep their identity for longer distances. 

We conclude this section by citing the work of Jackson \cite{Jakson}, where it can be noted that a possible design point for a kite made by a flexible membrane, like the ones available on the market, could be $C_L=0.55$ and $C_D=0.1$.
This optimized result has been obtained under several hypothesis: inviscid flow, lifting line theory (asymptotically accurate for large aspect ratios) and last, but not least, the need to maintain a constant tension in the kite canopy.
In our case, a rigid wing has been considered, showing that light but stiff structures, such as, for instance, inflatable wings, could be preferably employed to obtain a higher efficiency.
In fact, from the polar curve in figure \ref{2Dpolars}, it can be seen that the curved  wing at the same lift coefficient $C_L=0.55$ has a definitely lower drag coefficient $C_D=0.04$.
However, the design of similar structures requires the analysis of complex fluid structure interactions due to the small bending stiffness of the wing and the huge deformations that occur under aerodynamic loads. This kind of analysis was not the aim of the present work.

% \vspace{2mm}

\section{Conclusions}

In this work we have compared the aerodynamics of two rigid non twisted non tethered wings that are alike in all aspects (i.e. shape and profile section, aspect ratio and Reynolds number) but which differ in their curvature: an arc shaped curved wing which models a kite wing and reference straight wing. Given the lack of information on the transition on curved wings, we carried out a comparison between the flat and curved configuration by modelling the boundary layer on the wings as turbulent from the leading edge. %entire boundary
% In this work we have compared the aerodynamics of two rigid non twisted non tethered wings that are alike in all aspects, apart from the curvature (the same shape and profile section, the same aspect ratio and Reynolds number).$p_\infty = 101325$ Pa
The results were obtained through the computation of the numerical solutions of the Reynolds averaged Navier-Stokes equations (STAR-CCM+ code) for the mean flow.  We observed a slight deterioration of the overall aerodynamic performances of the curved wing (non-tethered kite) with respect to the flat configuration. Towards the wing tips, the lift on the curved wing was comparatively lower than that of the flat wing, due to a more extended separation region above the airfoil. % The separation region above the airfoil is in fact more extended in this area. 

A non trivial behaviour was observed in the vorticity dynamics in the near wake, up to five chords downstream from the trailing edge. %(up to a downstream distance equal to six chords from the leading edge).
The curved wing did not generate more intense wing tip vortices than the flat wing, however, the downstream decay in the near wake was slower. At the higher angle of incidence, $\alpha=18^{\rm o}$,  the curved wing induces a higher convergence of the wing ends-vortices. The flat wing instead maintained a constant convergence.
Such information could be useful for the design of a system configuration where a set of kites fly under mutual interference (in a ladder or carousel configuration), as proposed for wind generators systems.

% Given the  lack of information on the transition evolution on curved wings, we have carried out the comparison between the flat and curved configuration by setting the turbulent configuration of motion throughout the entire boundary layer on the wings.
% Given the lack of information on the transition evolution on curved wings, we have carried out the comparison between the flat and curved configuration by modelling the boundary layer on the wings as turbulent from the leading edge. %entire boundary layer as turbulent configuration of motion throughout the entire boundary layer on the wings.

Moreover, the data outlined in this study have three other implications. Firstly, they can be a first step for more advanced, unsteady simulations, namely large eddy simulation of the flow field near the wing and inside the wake. Furthermore, they could be used by designers of kites for wind power plants to improve the set up of automated non linear control systems. Lastly, they  could represent a basic mean flow to be used as an equilibrium starting point for perturbative stability analysis.

% The set of results here presented can be useful as a preliminary reference database for future, more advanced, unsteady numerical simulations, namely large eddy simulation of the flow field near the wing and inside the wake. Furthermore, the present data could be used by designers of kites for wind power plants to improve the set up of automated non linear control systems. Lastly, these results could represent a basic mean flow which could represent an equilibrium starting point for perturbative stability analysis. %, steady in the mean, frame for possible stability and optimization studies.

%Bibliografia

\end{document}